\documentclass[default,iicol]{sn-jnl}

\usepackage{xspace}
\usepackage{enumitem}
\usepackage{pdflscape}
\usepackage{afterpage}
\usepackage{capt-of}
\usepackage{graphicx}
\usepackage{url}

\newcommand\entityref[1]{\textsf{#1}}

\newcommand\ct[1]{\textsl{#1}}

\begin{document}

\title{Automatic Recommendations for Evolving Relational Databases Schema}

\author*[1]{\fnm{Anne} \sur{Etien}}\email{anne.etien@univ-lille.fr}

\author[1]{\fnm{Nicolas} \sur{Anquetil}}\email{nicolas.anquetil@univ-lille.fr}

\affil[1]{ \orgname{Univ. Lille, CNRS, Centrale Lille, Inria UMR 9189 - CRIStAL, Avenue Henri Poincaré, Villeneuve d'Ascq, 59655, France}}

\abstract{\emph{Context:} Relational databases play a central role in many information systems.
Their schema contains structural (e.g. tables and columns) and behavioral (e.g. stored procedures or views) entity descriptions.
Then, just like for ``normal'' software, changes in legislation, offered functionalities, or functional contexts, impose to evolve databases and their schemas.
But in some scenarios, it is not so easy to deconstruct a whished evolution of the schema into a precise sequence of operations.
Changing a database schema may impose manually dropping and recreating dependent entities, or manually searching for dependencies in stored procedures.
This is important because getting even the order of application of the operators can be difficult and have profound consequences.

This meta-model allows us to compute the impact of planned changes and recommend additional changes that will ensure that the RDBMS constraints are always verified.
The recommendations can then be compiled into a valid SQL patch actually updating the database schema in an orderly way.

We replicated a past evolution showing that, without detailed knowledge of the database, we could perform the same change in 75\% less time than the expert database architect.
We also exemplify the use of our approach on other planned changes.
}


\maketitle

\pagestyle{plain}

\section{Introduction}

Many industrial systems rely on relational databases (DB) to persist their data.
These databases are described by schemas that contain structural entity descriptions (e.g. tables, columns), and behavioral entity descriptions (e.g. stored procedures, views, triggers).
Structural and behavioral entities are referencing each other through foreign keys, function calls, or table/column references in queries.

Just as for programs, databases must continuously evolve \cite{Skou14a} to adapt to new requirements or changes in their running context.
But databases present significant differences with programs in the conditions of these evolutions:

First, Relational Database Management Systems (RDBMS) do not allow schema inconsistencies on some entities at any time during the evolution.
When renaming a method or a variable in a ``normal'' program, references to this method or variable can be temporarily invalid as long as they are corrected before compilation.
In an RDBMS, on the other hand, there is no edit and compile time, the database schema must be valid at all times.

Thus for example, if a column is a primary key referenced by a foreign key in another table, this column cannot be directly removed, the dependency between the foreign key and this primary key must first be removed.
Or if a column is referenced in a view, again it cannot be removed unless the view itself is first removed.

Because of this constraint, the database architect (DBA) must have, at all times, a complete map of all dependencies between the entities, to foresee all consequences (impacts) of a planned change.
``The task of evolving a relational database requires many complex operations
that must be carefully coordinated and sequenced in order to achieve the desired
state''~\cite{Schu21a}.
We found no tool that maintains such a map to help planning a change.
We propose an approach that will deconstruct a planned database schema change into a sequence of operations that will realize this change.

Another difficulty comes from stored procedures.
Because they can use different programming languages, their bodies are  often treated as unstructured text by the RDBMS such as PostgreSQL.
In the best case, the RDBMS will warn that a procedure makes a reference to a table or view, but it offers no assistance for changes impacting a stored procedure body.
In the worst case the RDBMS will not even be aware of the dependencies from the stored procedures, and their validity can only be checked at execution time.

Coming back to the example of removing a column, if this column is referenced in a stored procedure the RDBMS will allow the removal of the column, and an error will be raised later when the stored procedure is executed.
This is different from most programming situations, where the IDE, or at least the compiler,  would warn a programmer that a function or method is referencing a missing entity.

The literature schema evolution is abundant.
However, it mainly studies evolution of schemas only composed of tables and columns.
Most evolutions performed on a table or a column are not allowed by the RDBMS if they violate a functional constraint.
Consequently dealing with schema evolution in this context is simplified, and difficulties result in data management or the update of the program using the database.

There is also much research on the co-evolution of a database schema and the program that accesses it.
This falls outside the scope of this paper, we are not considering the external programs.

In a large study, Vassiliadis analyses the schema evolutions of 195 open-source projects~\cite{Vass21a}.
He reached the conclusion that \emph{schema evolution} is mostly absent from the typical Free Open Source Project [\ldots].
This absence is not due to the lack of its necessity but rather due to its difficulty.

In this paper, we propose an approach to deal with this difficulty by automatically computing all the induced and required operations to bring the schema in the desired state.
The approach is inspired by the notion of refactoring in programming environments~\cite{Robe96a, Robe99a} where a change is abstractly expressed (eg. renaming a method) and all the induced operations are automatically managed by the IDE (ie. renaming the method invocations).
Note however that the changes we are considering are not refactorings in the sense that they do not aim for behavior preservation, but rather to achieve a complex evolution requiring a number of atomic changes to be performed in an precise order.

We make ours the words of Peruma \emph{et al.} \cite{Peru21a} ``[program source code] are not the only software engineering artifacts that developers refactor in real-world systems. One such artifact is a database.''
Our goal is that the DBA can specify some desired operator(s) to be applied and the tool will decompose it in a sequence of operators that respects the constraints of the RDBMS.
Such operators include temporarily removing and then recreating entities.
The tool can also handle stored procedures code so that it is kept in sync with the state of the database schema.
When the DBA is satisfied with the proposed actions, the tool can generate an SQL patch implementing them in the proper order.

Our approach is based on a meta-model of dependencies between all entities in the database schema.
We have implemented and experimented our approach on a real-world PostgreSQL database, with actual required evolutions.
Our approach only considers the database schema and not the data. It is intended to propose a list of atomic steps that will achieve a complex evolution.

In this paper, we focused on the impact of evolutions on the database schema.
Saving and reloading the data when needed is left to the user.
However, this will often not be needed as many operations on the tables are handled by the DBMS which takes care of the data.
Our tool mostly focuses on the views and stored procedures which do not hold data themselves
.

This paper makes the following contributions:
\begin{itemize}
\item Proposing to apply software evolution mechanisms to database schema evolution;
\item A meta-model of dependencies between database schema entities;
\item Definition of change operators on these entities and how to compute the impact of the operator on the target entity and depending entities;
\item An algorithm to derive, from an initial operator, a list of additional operators to be applied to keep the database in a consistent state;
\item An algorithm to instantiate a valid SQL patch by merging and ordering all these operators.
\end{itemize}

This article is an extension of a paper published in CAiSE 2020 \cite{Delp20a}.
We added more operators to the tool, extended the description of the internals of the approach, including the generation of the SQL patch (merging and sorting of operators), and added new experiments.

This article is organized as follows, first, we present a brief overview of the approach (Section\ref{sec:overview}) based on meta-models to represent database schemas (Section \ref{sec:schema-modeling}) and possible evolutions of these database schemas (Section \ref{sec:evolution-modeling}).
Using these abstractions, we detail our approach to help an architect conceive all the steps of a database schema evolution (Section \ref{sec:approach}).
This approach was tested on three real evolutions that are presented and discussed (Section \ref{sec:validation}).
We close the paper by reviewing related work (Section \ref{sec:related}) and proposing our concluding remarks (Section \ref{sec:conclusion}).

\section{Overview of the approach}
\label{sec:overview}
Our approach aims to help the DBA to perform database schema evolutions, whatever these evolutions concern tables, columns, view, or stored procedures.
The main idea is to avoid the DBA to keep in mind all the relationships between the entities and to try to foresee the impact of a change on other entities. 
Our tool supports the DBA to keep the schema in a consistent state.
The DBA keeps control of all final changes performed on the schema.

\begin{figure}[htpb]
  \centering
  \includegraphics[width=\columnwidth]{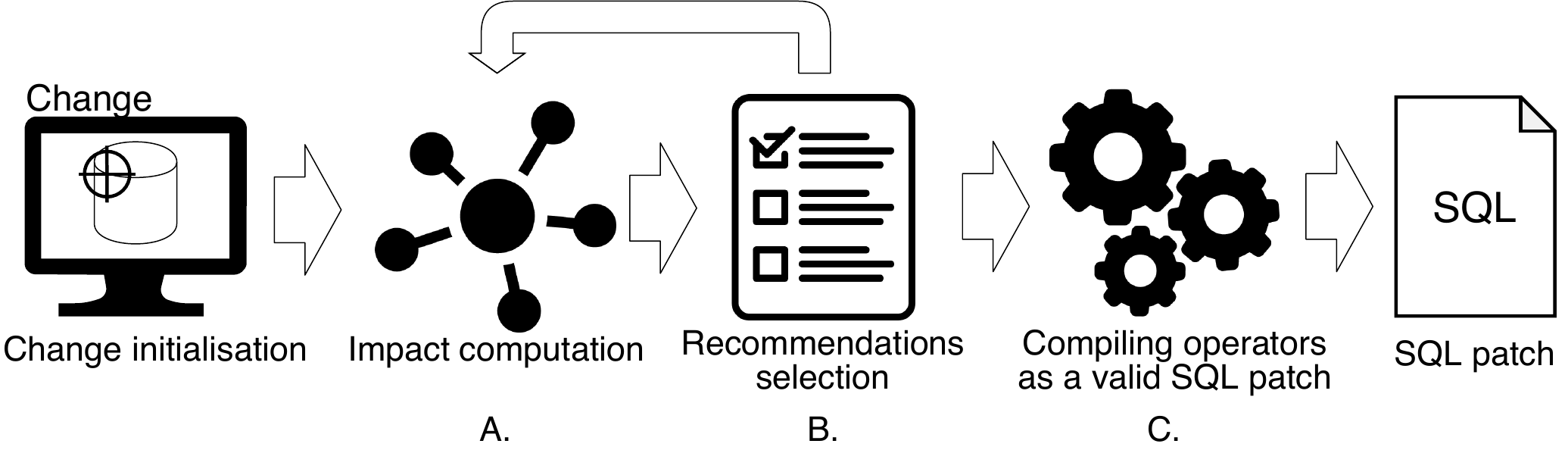}
  \caption{Coarse-grain illustration of the approach}
  \label{fig:process}
\end{figure}

We developed a 3-steps approach (see Figure \ref{fig:process}) working from an initial list of operators to apply.
This initial list (\emph{Change initialisation} in the figure) is given by the DBA according to the required evolution.
First (step \ct{A}), we compute the entities possibly impacted by the operator(s) in the list.
Then (step \ct{B}), we propose recommendation actions to mitigate this impact.
This second step can add new operators to the initial list.
Therefore, we iterate on steps \ct{A} and \ct{B} for all operators in the list (whether created by the user or added in step \ct{B}).
Once all impacts have been treated, we proceed to step \ct{C} where a SQL patch is generated from all the operators in the list.
The implementation of this approach is available on GitHub\footnote{\url{https://github.com/moosetechnology/DBEvolution.git}}.

Note that Bohnert and Arnold's definition of \emph{impact} \cite{Arno96b} mixes our steps \ct{A} and \ct{B}, that is to say computing the impacted entities and the actions to be performed to fix the introduced inconsistencies.
Their impact includes: 
``Identifying the potential consequences of a change, or estimating what needs to be modified to accomplish a change''.
We removed this confusion by having two separate steps.

Our approach relies on an abstract representation of the whole database schema. In the next sections, we first present the meta-models, and then we describe the evolutions that can be performed on it.

\section{Database Schema Modeling}
\label{sec:schema-modeling}

In this section, we present the database schema modeling, in the next one we will discuss how we model the possible evolutions of this schema.

We tried to make this meta-model as generic as possible.
This is eased by the normalized nature of relational databases and SQL.
However, each RDBMS vendor also introduces specificities that help it differentiate from the others.
In this paper, we concentrate on PostgreSQL databases until version 14.X. However, 
we believe adapting it to other RDBMS or newer PostgreSQL versions should remain a simple task.

The database schema meta-model has already been published in \cite{Delp20a}.
Its implementation is available on github\footnote{\url{https://github.com/juliendelplanque/FAMIXNGSQL}}.

\subsection{Schema Meta-Model: Overview}
\label{sec:schema-model-overview}

There can be two meanings for the term ``schema''.
In this paper, the concept of \emph{database schema} is used to refer to the way data are organized in a database (with tables, columns, views, constraints, \ldots).
It may also be called the \emph{database structure}.
This database schema can be obtained by performing an SQL dump of the database (disregarding the stored data itself).

Some RDBMSs also call ``schema'' a grouping of database objects.
For example in PostgreSQL, a schema is a namespace containing named objects (ex.: tables, views, stored procedures)

For us, this sort of ``schema'' is only one kind of database object that can be found in the ``database schema'' (first meaning) that we are modeling.

RDBMSs have allowed to store behavior inside the database (eg. stored procedure) for a long time.
This behavior is used, for example, to constrain data through triggers or \ct{CHECK} constraints.

In this article when the phrase \emph{database schema} (or \emph{schema}) is used, it refers to both structural and behavioral entities.
Because we are experimenting with PostgreSQL, when needing to refer to this second meaning (see Section~\ref{sec:schema-model-struct}), we will use the term ``namespace''.

Just like for ``normal code'', evolving a database schema implies modifying the entities composing the schema (tables, columns, \ldots) but also updating accordingly the references to these entities for example by taking into account functional dependencies.
For example, when renaming a table, a view referring to it must be updated; the same goes for stored procedures.
Our meta-model must model both schema entities and their references.
Schema entities are divided into structural and behavioural entities.
We, therefore, divide our database schema meta-model into three parts: (i) the structural part ie. mainly Table, Column, and Constraint (cf. Figure~\ref{fig:mm-structural}), (ii) the behavioural part corresponding mainly to views and stored procedures and thus also queries they embed (cf. Figure~\ref{fig:mm-behavioral}), and (iii) the cross-references between these entities representing the functional dependencies for example.

The meta-model is presented using UML convention with some slight additions:
inheritance links have straight corners while other links (associations) are rounded;
structural entities (ex: \entityref{Tables}) are red;
behavioral entities are orange (ex: \entityref{StoredProcedure});
and cross-references are white.

The meta-model presented here focuses on PostgreSQL because this is the RDBMS we used in our experiments.
Other RDBMS could have some additional concepts, but because Relational Databases mostly rely on the SQL standard, there should be little differences concerning the structure of the schema (Table, Column, and Constraint), the behavior (View and Stored Procedure) and the relationships between them.

\subsection{Schema Meta-Model: Structural Part}
\label{sec:schema-model-struct}
\begin{figure*}[htpb]
  \centering
  \includegraphics[width=\textwidth]{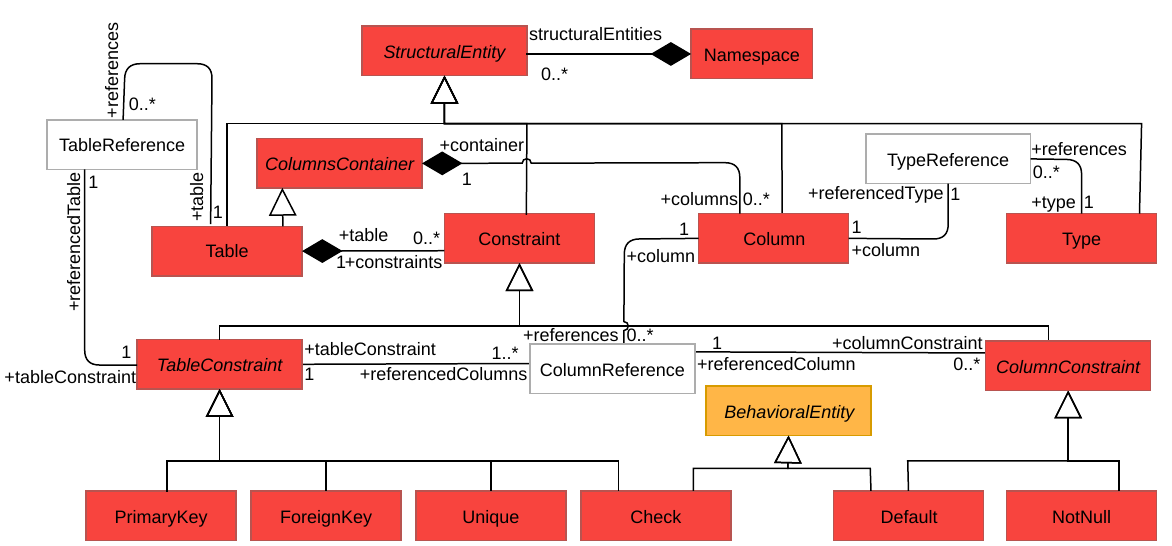}
  \caption{Structural entities of the meta-model.}
  \label{fig:mm-structural}
\end{figure*}

Figure~\ref{fig:mm-structural} shows the structural part of the meta-model.
A \entityref{StructuralEntity} defines the structure of data held by the
database or defining constraints applied to these data (eg. Table, Column, Referential integrity constraint, etc.).
The containment relation between \entityref{Table} and \entityref{Column} is
modeled through \entityref{ColumnsContainer} which is an abstract entity.
This entity also has sub-classes in the behavioral part of the meta-model (see Section~\ref{sec:schema-model-behavior}).
A \entityref{Column} has a type.
This relation is modeled through a \entityref{TypeReference}.
A \entityref{Column} can also be subject to \entityref{Constraint}s.
Depending on whether a \entityref{Constraint} concerns a single or multiple columns, it inherits from, respectively, \entityref{ColumnConstraint} or \entityref{TableConstraint}.
Six concrete constraints inherit from \entityref{Constraint}: \entityref{PrimaryKey}, \entityref{ForeignKey}, \entityref{Unique}, \entityref{Check} (a developer-defined constraint, described by a boolean expression), \entityref{NotNull}, and \entityref{Default} (a default value assigned when no value is explicitly provided, it can be a literal value or an expression to compute).
Note that \entityref{Check} and \entityref{Default} constraints also inherit from \entityref{BehavioralEntity} because they contain source code.
In addition, the concept of \entityref{Namespace} is a grouping of named database objects, somehow similar to packages in object-oriented languages.

\subsection{Schema Meta-Model: Behavioral Part}
\label{sec:schema-model-behavior}

\begin{figure*}[htpb]
  \centering
  \includegraphics[width=\textwidth]{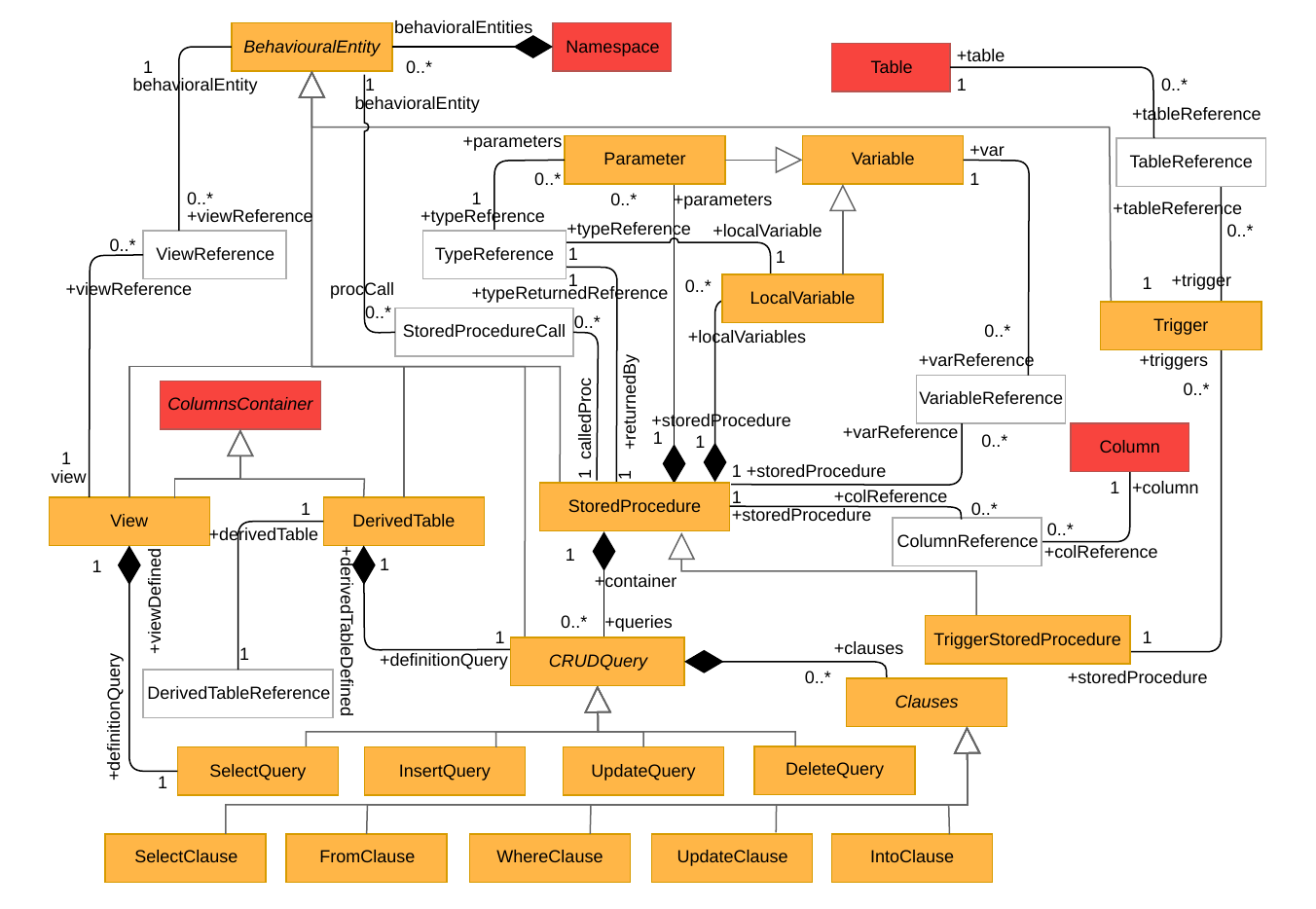}
  \caption{Behavioral entities of the meta-model}
  \label{fig:mm-behavioral}
\end{figure*}

A behavioral entity is an entity holding behavior that may act upon
\entityref{StructuralEntities}. Figure \ref{fig:mm-behavioral} shows
the behavioral part of the meta-model. The main entities are as follows.

\entityref{View} is a named entity holding a \ct{SELECT} query.
\entityref{StoredProcedure} is an entity holding developer-defined
behavior which includes queries and calls to other \entityref{StoredProcedure}.
We do not differentiate between functions and procedures.
A \entityref{StoredProcedure} contains \entityref{Parameter}(s) and
\entityref{LocalVariable}(s).
These entities can be referenced in clauses
of queries that are contained in \entityref{StoredProcedures}.
\entityref{Trigger} represents actions happening in response to an event on a
table (eg. row inserted, updated, or deleted).
Each subclass of \entityref{CRUDQuery} contains one or many clauses depending on
the query.
The possible clauses are: \entityref{With}, \entityref{Select},
\entityref{From}, \entityref{Where}, \entityref{Join}, \entityref{Union},
\entityref{Intersect}, \entityref{Except}, \entityref{GroupBy},
\entityref{OrderBy}, \entityref{Having}, \entityref{Limit}, \entityref{Offset},
\entityref{Fetch}, \entityref{Insert}, \entityref{Into}, \entityref{Returning},
\entityref{Update}, \entityref{Set}, \entityref{Delete}.
For the sake of readability, we did not include all the clause classes in the
diagram but only a few.
In a nutshell, the containment relation between CRUD queries and clauses is described in Table~\ref{tab:clauseContainment}.
\begin{table}[htb]
  \centering
  \caption{\label{tab:clauseContainment}Containment relations between CRUD
  queries and clauses.}
  \begin{tabular}{p{0.2\columnwidth}p{0.7\columnwidth}}
  \hline
  CRUD query & Clauses \\
  \hline
  \entityref{SelectQuery} &
  \entityref{With}, \entityref{Select}, \entityref{From}, \entityref{Where},
  \entityref{Join}, \entityref{Union}, \entityref{Intersect},
  \entityref{Except}, \entityref{GroupBy}, \entityref{OrderBy},
  \entityref{Having}, \entityref{Limit}, \entityref{Offset}, \entityref{Into}, \entityref{Fetch}\\

  \entityref{InsertQuery} &
  \entityref{With}, \entityref{Insert}, 
  \entityref{Returning}\\

  \entityref{UpdateQuery} &
  \entityref{With}, \entityref{Update}, \entityref{Set}, \entityref{From},
  \entityref{Where}, \entityref{Returning}\\

  \entityref{DeleteQuery} &
  \entityref{With}, \entityref{Delete}, \entityref{From}, \entityref{Where},
  \entityref{Returning}\\
  \hline
  \end{tabular}
\end{table}

\subsection{Schema Meta-Model: References}
\label{sec:schema-model-linking}

The third and last part of the meta-model represents links between entities.
It allows one to track relations between entities.
The references are fundamental in our approach. They help us to determine the impacted entities and thus consequently automatically propose to the DBA to avoid a dangling or out-of-date reference.
To simplify the approach, all references have been reified.
For example, a constraint will refer to a column through a \ct{ColumnReference} (Figures \ref{fig:mm-structural} and \ref{fig:mm-behavioral}). Thus, for example, a functional dependency is expressed through a \ct{ColumnReference} between a column and a \ct{ForeignKey} constraint. Similarly, a stored procedure accesses a local variable or a parameter through a \ct{VariableReference} and calls another stored procedure through a \ct{StoredProcedureCall}.

References are in white in Figures \ref{fig:mm-structural} and \ref{fig:mm-behavioral}.
In addition, Table \ref{tab:References} gathers the different entities that each clause of a
query may refer to.
For the sake of readability, those references were omitted in Figure~\ref{fig:mm-behavioral}.

Let us explain some non-obvious references that can appear in some clauses.
The first line of Table \ref{tab:References} specifies that a reference to a derived table can
appear in any clause.
This comes from the fact that a \texttt{SELECT} query can occur in any SQL
clause.
Line 2 comes from the fact that a \entityref{StoredProcedure} may
generate a table as a result, this is why it can appear in a \entityref{From}
clause.
Line 3 specifies that a reference to \entityref{Table} or \entityref{View} can
appear in clauses that normally deal with column references.
It comes from the fact that a developer can use a qualified reference to a
column (eg. \texttt{table\_name.column\_name}).
Finally, lines 3 and 4 specify that a stored procedure local variable or
parameter can appear in their respective set of clauses.
It occurs, for example, when the filter condition of a \texttt{WHERE} clause is
parameterized by a local variable or a parameter.
For example, \texttt{WHERE id = id\_to\_keep} where \texttt{id} would be a
reference to one of the columns of a table appearing in the \texttt{FROM} clause
of the query and \texttt{id\_to\_keep} would be a reference to a local variable
declared in the stored procedure holding the query.
\begin{table}[htbp]
    \centering
    \caption{\label{tab:References}References to entities from clauses.
    The other references to entities are shown in Figure~\ref{fig:mm-structural} and \ref{fig:mm-behavioral}}
    \begin{tabular}{cp{0.43\columnwidth}p{0.43\columnwidth}}
    \hline
    & Clauses & Entities \\
    \hline
    1 & \emph{any clause} & \entityref{DerivedTable}\\[1ex]

    2 & \entityref{From}, \entityref{Join},  \entityref{Into} &
    \entityref{StoredProcedure}, \entityref{Table}, \entityref{View}.\\[1ex]

    \raisebox{-0.5\baselineskip}{3} & \entityref{GroupBy}, \entityref{OrderBy}, \entityref{Having},
    \entityref{Set}, \entityref{Where}, \entityref{Returning},
    \entityref{Select}, \entityref{Update} &
    \entityref{Table}, \entityref{View}, \entityref{Column},
    \entityref{StoredProcedure}, \entityref{LocalVariable},
    \entityref{Parameter} \\[1ex]

    4 & \entityref{Limit}, \entityref{Fetch}, \entityref{Offset} &
    \entityref{StoredProcedure}, \entityref{LocalVariable}, \entityref{Parameter}\\
    \hline
  \end{tabular}
\end{table}

\subsection{Schema Meta-Model: Populating}
\label{sec:schema-model-building}

With this meta-model in hands, one must provide means to populate it.
We use a hybrid approach for this, where we first extract as much information as possible from the RDBMS meta-data to get a partial model of the database and then complete this model by analyzing the source code of behavioral entities.

The first part is RDBMS dependent as each will store meta-data in its own way.
However, it is relatively simple and usually well documented\footnote{System Catalogs description for PostgreSQL: \url{https://www.postgresql.org/docs/current/catalogs.html}}

The second step is language-dependent as different programming languages are typically allowed in stored procedures.
We developed an SQL parser to analyze view queries and a PL/pgSQL\footnote{PostgreSQL SQL Procedural Language: \url{https://www.postgresql.org/docs/current/plpgsql.html}} parser for stored procedures.

From the RDBMS meta-data, we instantiate a partial model that misses references made from behavioral entities.
It corresponds roughly to the meta-model in Figure~\ref{fig:mm-structural}.

From the RDBMS we also get the source code of each behavioral entity.
We apply a parser to analyze (statically) each string of source code.
This parser
(i) instantiates queries and clauses entities, and (ii) analyzes identifiers appearing in the
source code to create references in the model.

Incidentally, the source code of each entity is stored in the entity for later use (see Section~\ref{sec:approach-patch}).
For example, a \ct{SelectQuery} entity contains a string: \texttt{"SELECT}\ldots\texttt{INTO}\ldots \texttt{FROM}\ldots\texttt{"}.
Similarly, a stored procedure will contain the source code of its declaration.
For the references in the meta-model, we store their position (character offset) inside the SQL string of their owning entity.
For example, a \ct{ColumnReference} in a \ct{WhereClause} will store the positions of the first and last characters of the column's name in the query containing this clause.
This information is provided by the parser of the behavioral entities.

The source code of the meta-data reader\footnote{\url{https://github.com/olivierauverlot/PgMetadata}} and the parser\footnote{\url{https://github.com/juliendelplanque/PostgreSQLParser}} are available on GitHub.

\section{Database Schema Evolution Modeling}
\label{sec:evolution-modeling}

Having modeled the schema of a database, we now consider how to model the evolution of a schema. As a kind reminder, in this paper, we deal with evolutions of the database schema and the management of their impact on the rest of the schema to help the DBA to keep the whole schema in a consistent state. The management of the data during the application of the operators is not taken into account. 
Many evolutions are possible: eg. adding a table, removing a view, modifying a stored procedure.
We model them with \emph{operators}.
In this section, we list the operators considered by the approach, we review their implementation by RDBMS such as PostgreSQL, and we discuss issues arising when applying several operators to the same entity. 
Again, using another RDBMS could introduce other database objects and therefore other operators to add, remove, or modify them.
This is not a strong limitation as such new operators should be easy to similarly create to the ones we describe in this paper.

\subsection{Evolution Operators}
\label{sec:modeling-operators}

The database architect can express changes on the following entities of our structural meta-model (Section \ref{sec:schema-model-struct}):
\ct{Table, Column, Constraint, Type, View, Stored Procedure, Parameter, Trigger, Schema} and \ct{Derived Table}.
We do not consider ownership or security levels since we focus on structural changes that may have an impact on other entities.
We chose not to consider changes on \ct{Type} and \ct{Derived Table}.
We assumed that one very rarely changes a \ct{Type}.
That would be the case for example when changing a column's type \ct{float} with a constraint on the column that values should be positive, to a type \ct{positive float}. 
Similarly, derived tables are unusual, and changes concerning them are infrequent. 
Note that this is not a strong limitation and that these entities could be added to our approach.

Other entities are also not considered because they cannot be referenced \emph{per se} (\ct{Query}), or they have a small scope (\ct{Local Variable}).
In both cases, evolutions consist of simple changes of the containers (\ct{View} for \ct{Query} or \ct{Stored Procedure}; \ct{Stored Procedure} for \ct{Local Variable}) since the consequences are localized.
Thus, we model the evolutions of these entities as modifications of their container.

We consider the following \emph{evolution operators} on each of the remaining entities (see also Table~\ref{tab:operators}):
\begin{itemize}
 \item Add: this operator corresponds to the creation of an instance of the entity. It is mostly performed through a \ct{CREATE} SQL command.
 However, if the new entity is added to an, already existing, container, for example, a \ct{Column} added to a \ct{Table} or a \ct{View}, this can be done through an \ct{ALTER} or a \ct{CREATE OR REPLACE} SQL command.
 
 \item Rename: this operator aims to modify the name of an entity. 
 It is mostly performed with an \ct{ALTER} command.

 We chose to make a specific operator for this as the name often acts as an identifier for an entity in the RDBMS.

 \item Remove: this operator drops the entity from the database schema.
 When the entity is contained in another one (for example to remove a column or a constraint from a table), the removal might be performed through a modification of the container, either with an \ct{ALTER} SQL command or by dropping and recreating the container.

For example, removing a column \texttt{address} from an \texttt{EmployeeSimpleView} view is done by ``recreating'' the view without this column: ``\texttt{CREATE OR REPLACE VIEW EmployeeSimpleView AS SELECT employeeId, name FROM Employee;}''.

 \item Move: this operator aims to move an entity to another namespace.
 It would make no sense to apply it to entities contained in others.
 For example it would make no sense to try to move a column from one table to another one.
 Also, it would not make sense to move a trigger since they are always in the same namespace as the table to which they apply.

 \item Modify: this operator aims to modify the body of the behavioral entities or the type of a column in a table.
 In the first case, the operator is called \entityref{modifyBody}.
 It is performed with some of the many variations of the \ct{ALTER} commands. 
 The modification of the behavioral entity bodies is performed with a \ct{CREATE OR REPLACE} command.  
\end{itemize}
The principle of our approach is to help the DBA in the identification of the impacted entities and how to modify them to keep the schema consistent. 
Thus, the existence and the automatic management by PostgreSQL of the impact of an evolution operator on the referencing entities are discussed in section~\ref{sec:modeling-entity-dependencies}. Our approach helps the DBA to choose consistent changes resulting from the impact of the initial evolution operators. It is explained in section \ref{sec:approach}.

\afterpage{%
\begin{landscape}

\centering

\captionof{table}{Evolution operators for each entity kind with the associated SQL commands (only the initial part of the command is indicated).}
\label{tab:operators}
\begin{tabular}{lccccc}
\hline											
				&\textbf{Add}				&	\textbf{Rename}&	\textbf{Remove}& 	\textbf{Move}	&	\textbf{Modify}	\\
		\hline									
\small Table				& \small \ct{CREATE TABLE}				&	\small \ct{ALTER TABLE} 			&	\small \ct{DROP TABLE}		& \small	\ct{ALTER TABLE}	&	-	\\
\small Column				& \small \ct{ALTER TABLE}				&	\small \ct{ALTER TABLE} 			&	\small \ct{ALTER TABLE}		&	-	&	\small \ct{ALTER TABLE}	\\
\small Constraint			&  \small \ct{ALTER TABLE}				&	\small \ct{ALTER TABLE} 			&	\small \ct{ALTER TABLE}		&	-	&	\small \ct{ALTER TABLE}	\\
\small View				& \small \ct{CREATE VIEW}				&	\small \ct{ALTER VIEW}			&	\small \ct{DROP VIEW}		&	\small \ct{ALTER VIEW}	&	\small \ct{CREATE OR REPLACE}	\\
\small View Column			& \small \ct{CREATE OR REPLACE} 		&	\small \ct{ALTER VIEW}			&	\small \ct{DROP + CREATE}	&	-	&	-	\\
\small Stored Procedure		&   \small \ct{CREATE FUNCTION}			&	\small \ct{ALTER FUNCTION}		&	\small \ct{DROP FUNCTION}	&	\small \ct{ALTER FUNCTION}	&	\small \ct{CREATE OR REPLACE} 	\\				
\small Parameter			& \small \ct{CREATE OR REPLACE} 		&	\small \ct{CREATE OR REPLACE} 	&	\small \ct{DROP + CREATE}  	&	- & \small \ct{CREATE OR REPLACE} 	\\
\small Trigger				& \small \ct{CREATE TRIGGER}			&	\small \ct{ALTER TRIGGER}		&	\small \ct{DROP TRIGGER}	&	-	&	-	\\
\hline
\end{tabular} 

\end{landscape}
}

Table~\ref{tab:operators} sums up the different operators available for each entity and the associated SQL commands.
As mentioned, some evolution operators are not implemented in SQL and require to remove the entity, before re-creating it in the desired form.
This solution is only acceptable for entities that do not hold data (i.e. not tables), like views for example.

To give an idea of the level of details our operators allow, we list them in alphabetical order:
\begin{description}
\item[On columns:] \ct{Add}, \ct{Remove}, \ct{Rename}, \ct{Retype} operators;

\item[On constraints:] \ct{Add}, \ct{Remove} operators on \ct{CheckConstraint}, \ct{ForeignKeyConstraint}, \ct{NotNullConstraint}, \ct{PrimaryKeyConstraint}, and \ct{UniqueConstraint}.
\ct{Modify} operator on \ct{CheckConstraint};

\item[On stored procedures:] \ct{Add}, \ct{ModifyBody}, \ct{Move}, \ct{Remove}, \ct{Rename}, \ct{RenameLocalVariable}, \ct{RenameParameter}, \ct{RenameReferenceInStoredProcedure};

\item[On tables:] \ct{Add}, \ct{Move}, \ct{Remove}, \ct{Rename} operators;

\item[On triggers:] \ct{Add}, \ct{Remove}, \ct{Modify} operators;

\item[On views:] \ct{Add}, \ct{Remove}, \ct{Move}, \ct{ModifyBody}, \ct{RenameReferenceInNonSelectClause}, \ct{RenameReferenceInSelectClause}

\end{description}

\subsection{Special Operators}
\label{sec:modeling-special-operators}

For our approach, we also need to have some special operators:
\begin{description}
\item[Identity operator:] This is an operator that keeps an entity exactly as it is, but it does so by removing the entity and recreating it in the same form.
We will see that this is sometimes necessary, for example, some changes on a table that is referenced by a view require to remove the view first.
After the change on the table, the view is recreated as it was before the change.

\item[\ct{DoNothing}:] As the name suggests this operator does not make any change to the database schema but is useful to handle the consequences of some operators on referencing entities.

\item[\ct{HumanDecision}:] This operator is used, again to handle the consequences of applying a first operation, when no automated decision can be taken.
For example, if we remove a stored procedure that is called by another stored procedure, we cannot know what the DBA intends to do with the second one and this wild card operator is used to indicate that he must handle manually the consequences.

\item[Reference oriented operators:] 
SQL commands are traditionally \emph{entity oriented}, they operate on entities like tables, views, columns, constraints, {\ldots}
For convenience, we also have \emph{reference oriented} operators.
For example, we added operators like \ct{RenameReferenceInConstraint}, \ct{RenameReferenceInSelectClause}, \ct{RenameReferenceInNonSelectClause}, or \ct{RenameReferenceInStoredProcedure}.
In SQL, such reference-oriented operations are performed by modifying the entity making the reference for example with the SQL commands \ct{ALTER} or \ct{CREATE OR REPLACE}.
Being more specific on the kind of modification that needs to be performed allows us to handle it automatically.
\end{description}

\subsection{Actionable entities}
\label{sec:modeling-entity-actionable}

As we just noted, we created \emph{reference oriented operators} to finely represent the changes to be performed on a database schema.
However, these changes will be actually performed by SQL commands that are traditionally \emph{entity oriented} (they act on tables, views, constraints, etc.)

To convert our \emph{reference oriented operators} to \emph{entity oriented} ones\footnote{This is described in Section~\ref{sec:approach-patch}}, we will need to find what is the \emph{actionable entity} of an operator.
The \emph{actionable entity} is the containing SQL entity on which we can apply an SQL command to achieve the goal of an operator.

For example, to change a clause of the \ct{SelectQuery} of a view, the actionable entity is the view.
It can be changed by an \ct{ALTER VIEW} or a \ct{CREATE OR REPLACE} SQL command (see Table~\ref{tab:operators}).

For operators that are already \emph{entity oriented} (ex: \ct{RemoveTable} or \ct{RenameStoredProcedure}), the actionable entity is the target of the operator.
For reference-oriented operators,  the actionable entity can be found by going up the containment tree of entities (see also the meta-model in Figures \ref{fig:mm-structural} and \ref{fig:mm-behavioral}): a \ct{SelectClause} belongs to a \ct{SelectQuery} which, for example, belongs to a \ct{View}.

\subsection{Dependencies between entities}
\label{sec:modeling-entity-dependencies}

To deal with dependencies in a database schema, many changes are implemented in SQL by the \ct{ALTER} or \ct{CREATE OR REPLACE} commands.
Some schema evolutions are refactorings that are part of the SQL language. 
For example, when renaming a table (with \ct{ALTER TABLE}) referenced through a foreign key constraint, the reference is automatically updated, keeping the database in a consistent state.
However, not all such induced evolutions are automatically managed.
As can be seen in Table~\ref{tab:operators}, renaming a column of a view ie. appearing in the \ct{SELECT} clause of the view must be done by removing (\ct{DROP}) the view and recreating it.
In the preceding section, we mentioned the fact that there are no command in SQL to change the table referenced by a view, the entire body of the view has to be redefined.

Table \ref{tab:entity-dependence} summarizes the possibility of applying a given evolution operator (first column) on a referenced entity (second column) depending on what other entity references it (right part of the table).
We do not consider the \ct{Add} operator in this table because new entities cannot be already referenced (except by Stored Procedures which are unchecked).
We do not consider triggers and constraints either because they cannot be referenced by another entity. 
The \ct{Modify} operators are not considered because either the modification changes the way an entity is identified (name, signature, or name of column for views) and it is already considered through other operators, or it changes the body of a behavioral entity and has no consequence for referencing entities.

There are three possible outcomes:
\begin{description}
\item[auto:] The RDBMS allows to perform the operation (see SQL commands in the previous section) and automatically (and silently) updates the referencing entity (for example renaming a table referenced by a view will also change the view's query);

\item[no:] The RDBMS does not allow us to perform the change to avoid leaving the database in an inconsistent state.
A common solution to this is to remove all references to the target entity.
For example, to remove a table referenced by a view one must either change the query of the view (to remove references to the table) or remove the view altogether.

\item[unchecked:] (for stored procedure) The RDBMS will perform the change without any check.
This can leave dangling references in the stored procedure that will only be detected on the next execution of this procedure by raising an error.

\end{description}

\begin{table*}
\begin{center}
\caption{Availability of evolution operators when the target entity is referenced (``no'': forbidden; ``auto'': automatic; ``unchecked'': allowed (not checked); "-": operator  makes no sense)}
\label{tab:entity-dependence}
\begin{tabular}{cc|cccc}
\hline											
\textbf{Evolution} &	\textbf{Changed}	&	\multicolumn{4}{c}{\textbf{Referencing entity}}					\\		
\textbf{operator}&	\textbf{entity}	&	\textbf{Table}	&	\textbf{View}	&	\textbf{Stored Proc.}	&	\textbf{Trigger}	\\
\hline
Rename	&		&	auto&	auto	&	unchecked	&	auto	\\
Remove	&	Table	&	no	&	no	&	unchecked	&	no	\\
Move	&		&	auto&	auto	&	unchecked	&	auto\\
\hline
Rename	&	&	auto&	auto	&	unchecked	&	auto	\\
Remove	& \raisebox{0pt}[0pt][0pt]{\parbox{5em}{\centering Table column}} &	no	&	no	&	unchecked	&	no	\\
Move	&		&	-	&	-	&	-	&	-	\\
\hline
Rename	&		&	-	&	auto 	&	unchecked	&	-	\\
Remove	& View	&	-	&	no	&	unchecked	&	-	\\
Move	&		&	-	&	auto	&	unchecked	&	-	\\
\hline
Rename	&		&	-	&	no	&	unchecked	&	-	\\
Remove	& \raisebox{0pt}[0pt][0pt]{\parbox{5em}{\centering View column}} &	-	&	no	&	unchecked	&	-	\\
Move	&		&	-	&	-	&	-	&	-	\\
\hline
Rename	&		&	-	&	auto	&	unchecked	&	auto	\\
Remove	& \raisebox{0pt}[0pt][0pt]{\parbox{5em}{\centering Stored procedure}} &	-	&	no	&	unchecked	&	no	\\
Move	&		&	-	&	auto	&	unchecked	&	auto	\\
\hline											
\end{tabular} 
\end{center}
\end{table*}

There is a real challenge to provide a tooled assistance for the two latter cases. 
When the RDBMS does not allow an operation, we will remove the road blocks preventing it, for example by removing referencing entities.
After the operation is performed, we will automatically recreate the referencing entities as they were before (\emph{Identity operator}, Section \ref{sec:modeling-operators}).
When the RDBMS allows an operation without doing any checks (stored procedures), we will perform the checks and make the needed adjustments so that the procedure is still valid.
The DBA should be able to just state the change he has in mind and the tool semi-automatically handles the consequences on dependent entities. 
This is all the more difficult when several changes need to be made on the same entity (see Section~\ref{sec:patch-merge}).
This can happen because one change has several consequences, or because the DBA requires several changes all impacting the same entity.



\section{Description of the Evolution Approach}
\label{sec:approach}

Our approach works on a model of the database schema.
This allows one to temporarily relax schema constraints and dangling reference constraints on the model for the sake of evolution.
It allows the developer to focus on changes to be made and not on how to fulfill schema consistency constraints and avoid dangling references at any time.
These considerations are handled by our tool which detects the inconsistencies that an evolution would threaten to introduce and proposes actions to resolve them.

\begin{figure}[htbp]
  \centering
  \includegraphics[width=0.95\columnwidth]{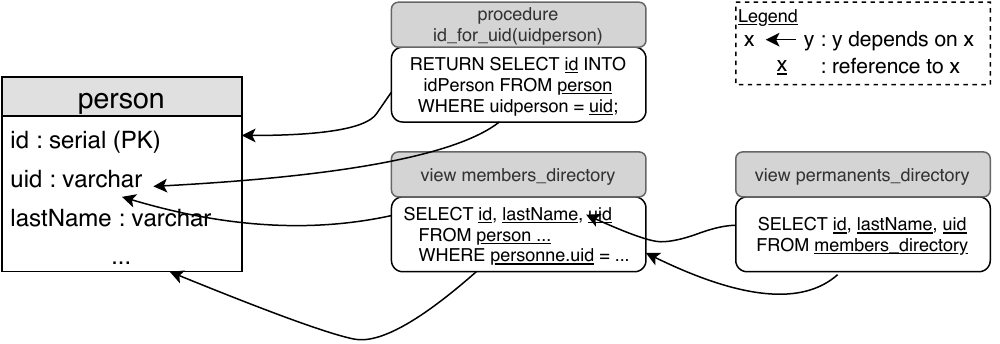}
  \caption{A small example of dependencies between a table, two views, and a stored procedure}
  \label{fig:exampledb}
\end{figure}

We will illustrate the proposed process with a simplified example from a real evolution \cite{Delp18c}.
Figure \ref{fig:exampledb} represents a table \ct{person}, two views \ct{members\_directory} and \ct{permanents\_directory}, and a stored procedure \ct{id\_for\_uid()}.
Dependencies between entities are modeled with arrows.
For example, the arrow between \ct{id\_for\_uid()} and \ct{person} is an instance of TableReference and the arrow between \ct{id\_for\_uid()} and \ct{uid} is an instance of ColumnReference.

Column \ct{person}.\ct{id} is a serial corresponding to the primary key of table \ct{person};
\ct{uid} is a varchar referring to the User ID of a person in an external LDAP (Lightweight Directory Access Protocol);
\ct{lastName} is a varchar.
The planned evolution consists of renaming the \ct{person}.\ct{uid} column into \ct{person}.\ct{login}.

\subsection{Change Impact}
\label{sec:approach-impact}

In our example, the initial list of operator(s) will contain a \ct{RenameColumn} operator on \ct{person.uid}.
This represents the evolution to perform.

The \emph{potential impact} of changing an entity is defined as the set of all entities that refer to this entity.

For example, the potential impact of the \ct{RenameColumn} is all the \entityref{ColumnReference}s on column \ct{uid} of table \ct{person}: 
\begin{enumerate}

\renewcommand{\theenumi}{(\roman{enumi})}%
\renewcommand{\labelenumi}{\theenumi}%
\item in the query of the stored procedure \ct{id\_for\_uid()};
\item in the \ct{SELECT} clause of the query defining view \ct{members\_directory}; and
\item in the \ct{WHERE} clause of the query defining view \ct{members\_directory}.
\end{enumerate}

These three clauses (entities in our meta-model, see Section \ref{sec:schema-model-behavior}) are added to the \emph{potential impact} of renaming \ct{person.uid} as \ct{person.login}.

To ease the \emph{recommendation} step, the set of potentially impacted entities is split into disjoint ``coherent subsets'' of entities.
For example, \ct{TableReferences} can originate from SQL queries, constraints, or triggers.
If entities of these three types are in the potential impact set, they will be separated into three subsets, one for each type.
The definition of what is a ``coherent subset'' of entities depends on the operator.
For example, we might want to treat differently a \ct{ColumnReference} in the \ct{SELECT} clause from one in the \ct{WHERE} clause (two separate ``coherent subsets''), or it might be possible to treat them the same way (only on ``coherent subset'').
These subsets depend on the recommendations that we want to propose in the next step.
The only rule is that, for a given operator, all ``coherent subsets'' of entities must form a partition of the potential impact set (i.e. all potentially impacted entities must belong to exactly one subset).

The identification of all the possible coherent subsets of impacted entities relies first on the identification of the possibly impacted entities (the ones referencing the changed entity), and second on a good understanding of how SQL databases work.
It is intended to be defined by the developers of the tool (here the authors) and not by its users.

All possible types of referencing entities (set of impacted entities) are deduced from the meta-model (Figures~\ref{fig:mm-structural} and \ref{fig:mm-behavioral}) and Table~\ref{tab:References}.

For example, in the case of the \ct{RenameColumn} operator acting on a \ct{Column} (from a \ct{ColumnContainer}), one must understand all possible \ct{ColumnReferences}.
From Figure~\ref{fig:mm-structural} we find that two entity types may have a \ct{ColumnReference}: \ct{TableConstraint} and \ct{ColumnConstraint}.
From Figure~\ref{fig:mm-behavioral}, we add the entity type \ct{StoredProcedure}.
Finally, from Table~\ref{tab:References}, we see all the clauses of a \ct{CRUDQuery} that can reference a \ct{Column} (on the third line of the table), and going back to the meta-model (Figure~\ref{fig:mm-behavioral}) we see what entities can contain a \ct{CRUDQuery} and therefore make a \ct{ColumnReferences}: \ct{StoredProcedure} again, \ct{DerivedTable}, and \ct{View}.

From this, and knowing, for example, the differences between a column reference in the \ct{Select} clause of a \ct{View} (giving the column of the \ct{View}) and a column reference in another clause of the same \ct{View} (no external impact for the schema), one deduces the following ``coherent subsets'' for the \ct{RenameColumn} operator:
\begin{enumerate}

\renewcommand{\theenumi}{(\roman{enumi})}%
\renewcommand{\labelenumi}{\theenumi}%
\item \label{item:subset-constraint}
	all the constraints that refer to the column; 
\item \label{item:subset-select}
	all the select clauses of view queries referring explicitly to the column (thus, not wildcard ``\ct{*}''); 
\item \label{item:subset-where}
	all the other clauses of view queries; 
\item \label{item:subset-procedure}
	all the clauses of queries embedded in stored procedures.
\end{enumerate}

Cases \ref{item:subset-select} and \ref{item:subset-where} are different because changing the selected columns of a view (case \ref{item:subset-select}) requires to drop and recreate the view (see Table~\ref{tab:operators}) while modifying the ``body'' of the view query (case \ref{item:subset-where}, eg. the \ct{WHERE} clause,) is done with a \ct{CREATE OR REPLACE VIEW} command.
Since the consequences of the change are different, they are in separate subsets.

For each evolution operator, one must identify beforehand all the possible coherent subsets of impacted entities.
This is done by the developer of the tool as explained above for \ct{RenameColumn} and remains valid as long as the semantics of the operators do not change.

\subsection{Recommendations}
\label{sec:approach-recommendation}

Once the \emph{potential impact} of a change is computed, decisions might need to be made to handle impacted entities.
We call each of these decisions a \emph{recommendation}.

Associated to each coherent subset of impacted entities, one must also identify the consequences of the operator (on the impacted entities of this subset), and the recommendation(s) to handle potential inconsistencies that the operator could raise.
For example, for the \ct{RemoveColumn} operator, our approach will recommend to remove possible constraints on columns referencing it (\ct{Foreign Key}).

When there are several possible recommendations for one coherent subset of impacted entities, they are all proposed to the user (the DBA) that chooses for each potentially impacted entity which recommendation to apply.
Currently, the recommendations are proposed to the user in no particular order.
A future improvement could be to define a meaningful order (see Section~\ref{sec:conclusion}).
These choices mainly correspond either to the will to propagate the change, as much as possible, to reflect the intention of the DBA or in the opposite to limit it, for example by aliasing a column of a view whose name in the original entity has changed to not create a new change in this view.
This choice creates a new evolution operator that is added to the initial list of operators to apply.
We then loop on steps \ct{A} and \ct{B} to identify the potential impact of these new operators and the associated recommendations.

In our running example, for each of the three potentially impacted entities, \emph{recommendations} are produced:
\begin{itemize}
\item For the query in the stored procedure \ct{id\_for\_uid()}, the recommendation is to replace the reference to column \ct{uid} with a reference to column \ct{login}.
This is materialized as an operator \ct{RenameReferenceInStoredProcedure}, added to the initial list of operators to apply.

Although the change seems obvious, it is important to model it explicitly because it could create new impacts.
This is not the case in our example.

\item For the reference to \ct{uid} in the \ct{WHERE} clause of view \ct{members\_directory}, the recommendation is again to replace the reference to column \ct{uid} with a reference to column \ct{login}.
An operator \ct{RenameReferenceInNonSelectClause} is added to the initial list of operators.
Again, the potential impact of this new operator is computed in another iteration and proves to be empty because changing the \ct{WHERE} clause does not change the ``format'' of the query result (the result has the same columns).

\item Finally, for the reference to \ct{uid} in the \ct{SELECT} clause of view \ct{members\_directory}, two recommendations are proposed to the architect:
  \begin{itemize}
  \item aliasing the column (ie. replacing ``\ct{SELECT uid}'' by ``\ct{SELECT login AS uid}'') or,
  \item as before, replacing the reference to \ct{uid} by a reference to \ct{login}.
  \end{itemize}
If the DBA chooses the latter case, the operator \ct{RenameReferenceInSelectClause} is added to the initial list of operators. Changing the column \ct{uid} in view \ct{members\_directory} to \ct{login} has a potential impact on the view \ct{permanents\_directory} (see again Figure \ref{fig:exampledb}).
This impact is detected and handled in a new iteration of steps \ct{A} and \ct{B}.

\end{itemize}

\begin{figure*}[htbp]
  \includegraphics[width=0.97\textwidth]{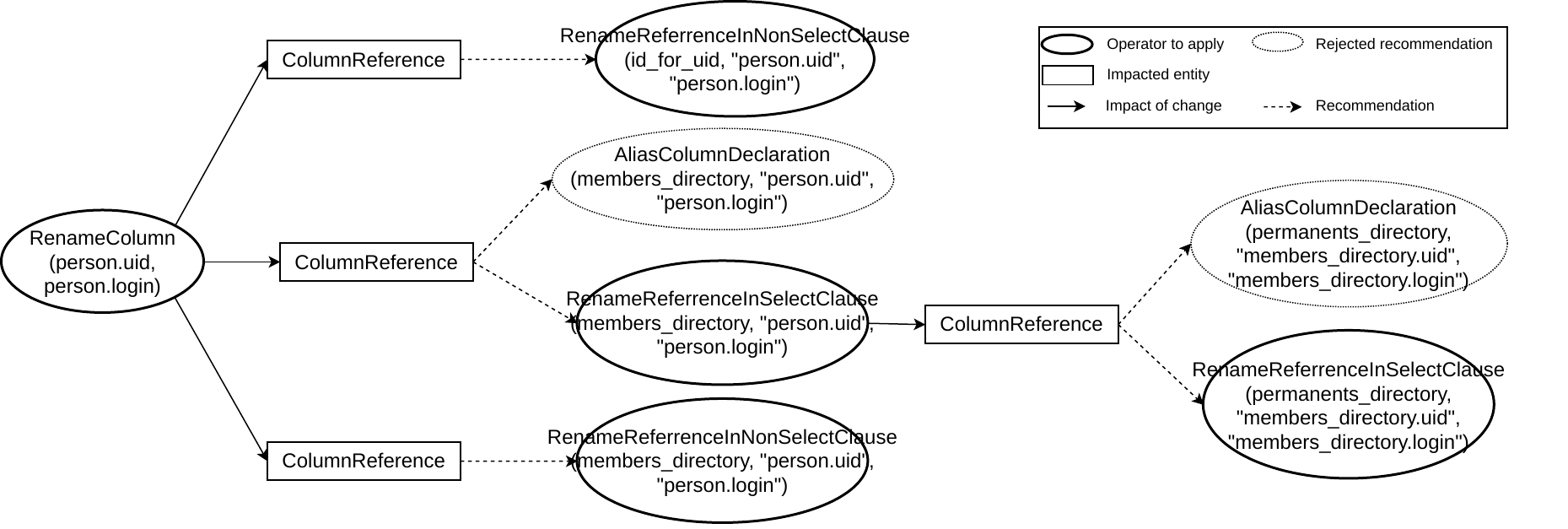}
  \caption{Recommendations selection for renaming \ct{person.uid} on the example of Figure \ref{fig:exampledb}}
  \label{fig:decisionProcess}
\end{figure*}

The final output of applying all steps \ct{A} and \ct{B} is a list of hierarchical trees of operators where the root of each tree is one of the operators the DBA originally specified (what he wanted to do) and the inner nodes and leaf of the trees are operators that were added to handle impacts.
Figure \ref{fig:decisionProcess} illustrates the tree of operators for our running example.

\subsection{Generating a Valid SQL Patch}
\label{sec:approach-patch}

When the hierarchical tree of operators is complete for all the operators initially stated by the DBA, we can proceed to step \ct{C}: 
Generating a SQL patch that will apply all these operators in an orderly fashion.
This means for example that all operations marked ``no'' in Table \ref{tab:entity-dependence} are treated to eliminate the reference(s) that could prevent their application.
Just as a DBA would normally do, we treat this by removing the references, performing the operation, and then recreating the references at the end of the patch so that no entity (table, view, constraint, \ldots) is lost unless the DBA explicitly decided so.

Schuler and Kesselman \cite{Schu21a} state that ``The task of evolving a relational database requires many complex operations that must be carefully coordinated and sequenced in order to achieve the desired state.''
To ensure that the patch executes flawlessly, we must take several precautions:
\begin{itemize}
\item for convenience, we used \emph{reference oriented} operators (for example \ct{RenameReferenceInSelectClause}, see Section \ref{sec:modeling-operators}).
They need to be translated into SQL commands which are typically \emph{entity oriented}.

\item identity operators leave an entity unchanged by removing it and recreating it as it was.
These identity operators need to be decomposed in two corresponding SQL commands \ct{DROP} and \ct{CREATE}.
These two commands need to be ordered appropriately in the SQL patch (see last point);

\item the final SQL patch might contain several commands targeting the same SQL entity (e.g. several \ct{ALTER VIEW} changing different aspects of a view's query), they all need to be merged into one single SQL command so that some changes do not override previous ones.

\item the operations to perform must be ordered so that, when necessary (see Table  \ref{tab:entity-dependence}) references to entities are removed before the entities are changed, and referred entities are created before references to them are introduced.
\end{itemize}

We now describe the actions to ensure these precautions are properly taken.

\subsubsection{Reference oriented operations}

\label{sec:patch-ref-oriented}

We created reference-oriented operators, such as \ct{RenameReferenceInSelectClause}, that are more specific than the available SQL commands.
These operators need to be translated into actual SQL commands.
For this we need to find the \emph{actionable entity} to change, as described in Section~\ref{sec:modeling-entity-actionable}.
(The actionable entity is the entity on which to apply the SQL command.)

Each reference-oriented operator ``knows'' how to be translated according to the actionable entities it may appear in.
This is handled through the coherent subsets of impacted entities (see Section~\ref{sec:approach-impact}).
In the case of renaming a column, the operator has different coherent subsets whether its parent query belongs to a view or a stored procedure.

To help generating the proper SQL command we use the SQL source code stored in the entities of the model (see Section~\ref{sec:schema-model-building}).
Thus, the \ct{SelectQuery} entity in our stored procedure (\ct{id\_for\_uid()}) contains the string: \texttt{"SELECT id INTO idPerson FROM person}{\ldots}\texttt{"}, and the \ct{ColumnReference} to \ct{person.uid}) stores its starting and ending position (character offset) inside this SQL string
It is therefore a simple matter of substring substitution to change a reference in a query from one name to another and to re-generate the proper SQL code.

\subsubsection{Identity operations}

\label{sec:patch-identity}

We already saw that some commands may not be applied if there is a reference on the targeted entity.
The referring entity does not need to be modified, but it must be removed before changing the target entity.
To deal with these cases, we created an identity operator (section \ref{sec:modeling-operators}) that will remove an entity and recreate it later exactly as it was before.

This operator trivially translates to a \ct{DROP} and a \ct{CREATE} SQL commands.
Note that this may be a recursive process where to remove a referring entity, one must first remove another entity itself referring to it.
Finding dependencies is easy as they are all modeled (white boxes in figures \ref{fig:mm-structural} and \ref{fig:mm-behavioral}). 
As in the previous point, this is based on the actionable entities that are the ones to drop and recreate.
The only point to take care of, is the order in which the removal and recreation are performed.
This is discussed in the last point (sub-section \ref{sec:patch-order}).

\subsubsection{Merging operations}

\label{sec:patch-merge}

Several operators may target the same SQL entity either because one change has several consequences or when several changes requested by the user impact the same entity.
For example, when several references need to be renamed in the body of a behavioral entity, or when a move and a rename operator are applied to the same entity.
In this case, if we generate SQL code for each operator independently the result will not be as expected.
For example, the name of a table acts as an identifier for it.
Therefore if moving and renaming a table (two successive operators), the second operator would need to take the effect of the first into account to work.

All operators on one given SQL entity need to be merged in one single SQL command.
This merge is done by comparing the actionable entity of each operator.
If two operators have the same actionable entity, we merge their effects in one single SQL command.

We may also remove duplication or simplify the patch here.
For example in a case where one operator removes an entity that another operator was modifying, the second one becomes irrelevant.

One could also deal with contradictions, for example, if one operator renames a column to A and another renames the same column to B.
These contradictions would have been introduced by the DBA when dealing with individual recommendations (Section \ref{sec:approach-recommendation}).
We do not check for contradictions at this point, but a simple solution would be to signal them and stop the patch generation step.

\subsubsection{Ordering operations}

\label{sec:patch-order}
This step is also challenging since the generated SQL patch must ensure both that all the changes provided by the DBA are performed and that the database is always in a correct state.
Finally, when all the SQL commands are generated, they must be ordered in an SQL patch.
For this, we use the dependencies between the entities targeted by the SQL commands.
These references form a, possibly disconnected, graph of dependencies.
The graph is directed and may not contain any cycle.
for example, two views cannot mutually reference each other, or if two tables declare foreign keys on each other primary key, the dependencies are from the \ct{ForeignKeyConstraints}, not between the tables themselves.

The patch begins with all \ct{Remove} commands.
These commands might come from the first part of identity operators (\ct{DROP} + re-\ct{CREATE}), or a need of the DBA.
They are applied in order of dependencies: if an entity A depends on (i.e. references) an entity B, then A is removed before B.
Because there can be no cycles in the dependency graph, a total order of the commands can always be created.

Once this is done, other SQL commands are applicable, even those marked ``no'' in Table \ref{tab:entity-dependence} because we removed the references.
All remaining operators are inserted in the patch, again in reverse order of dependencies: if an entity A depends on (i.e. references) an entity B, then B is modified/created before A.
This includes the \ct{Create} operators coming from the second part of our identity operator.

For security, the patch is created inside a \ct{BEGIN} / \ct{ROLLBACK} transaction and presented to the DBA.
He may then review it, and if satisfied, copy it into the RDBMS client, and \ct{COMMIT} it.

The final patch generated for our example is listed in Figure \ref{fig:example-patch}.
Note that:
\begin{itemize}
\item The \ct{RenameReferenceInSelectClause} for the query of view  \ct{permanents\_directory} is translated to a \ct{DROP} command at the start of the SQL patch and a \ct{CREATE} command at the end.
This is necessary to change the SELECT query of the view;

\item A similar thing happens for the query of view \ct{members\_directory}, but additionally, two operators (\ct{RenameReferenceInSelectClause} and \ct{RenameReferenceInNonSelectClause}) were merged in the \ct{CREATE} command;

\item the \ct{CREATE OR REPLACE FUNCTION} command could occur anywhere in the patch since its references are not checked by the RDBMS;

\item the operation initially required by the DBA (\ct{RenameColumn} \ct{person.uid}) appears in the middle of the patch, with all \ct{DROP} commands before it and all \ct{CREATE} commands after it.
This is the usual behavior, that drops the references before the required operation and recreates them afterward.
\end{itemize}

\begin{figure}
\begin{center}
\begin{small}
\begin{verbatim}
BEGIN;
DROP VIEW "permanents_directory" RESTRICT;
DROP VIEW "members_directory" RESTRICT;

CREATE OR REPLACE FUNCTION
  "id_for_uid"(uidperson varchar)
  RETURNS int4 AS $$
DECLARE
  idperson int4;
BEGIN
  SELECT id INTO idperson
  FROM
    person
  WHERE
    uidperson = login;
  RETURN idperson;
END;$$ LANGUAGE plpgsql;

ALTER TABLE "person"
  RENAME COLUMN "uid" TO "login";

CREATE VIEW "members_directory" AS
  SELECT
    person.id,
    person.lastname,
    person.login
  FROM person
  WHERE ((person.login)::text = ...);

CREATE VIEW "permanents_directory" AS
  SELECT
    members_directory.id,
    members_directory.lastname,
    members_directory.login
  FROM members_directory;
ROLLBACK;
\end{verbatim}
\end{small}
\caption{SQL patch produced for our running example}
\label{fig:example-patch}
\end{center}
\end{figure}

\subsection{A tool to support database schema evolution}
\label{sec:approach-tool}

\begin{figure}[htb]
  \centering
  \includegraphics[width=\columnwidth]{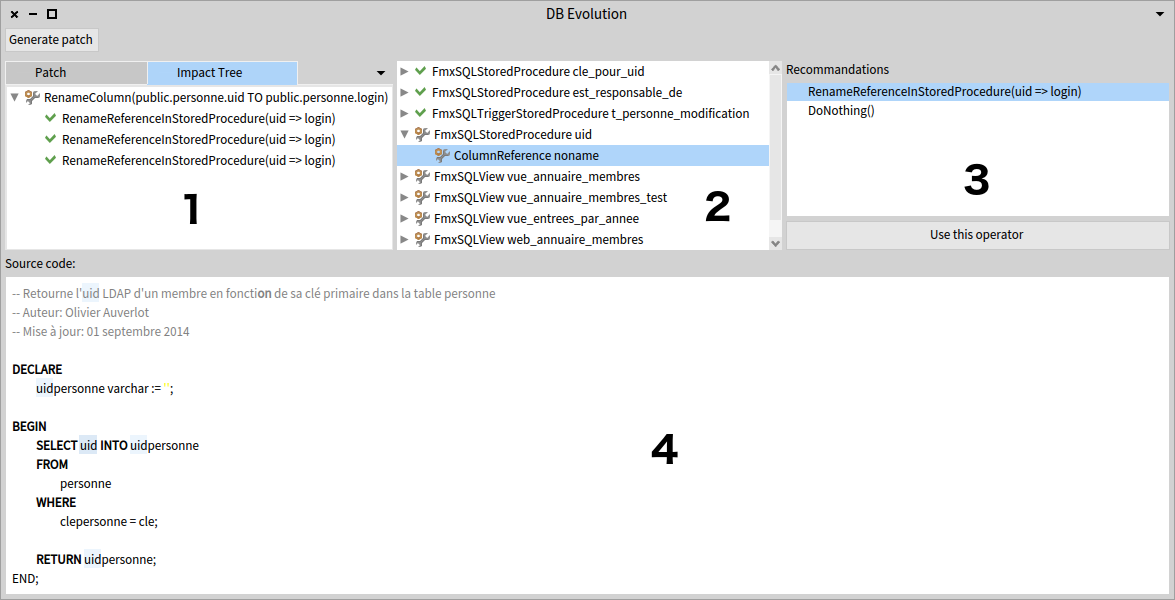}
  \caption{Screenshot of DBEvolution, the implementation of our approach that we
  used to perform the experiment.}
  \label{fig:dbevolutiontool}
\end{figure}

We implemented a graphical user interface for our approach.
Figure \ref{fig:dbevolutiontool} shows a screenshot of DBEvolution graphical user interface.
It guides the user through the steps of choosing recommendations.

Panel~1 shows the list of operators selected by the user and the tree of impacts resulting from the user's choices.
When an operator is inserted or clicked in panel~1, panel~2 shows the actionable entities potentially impacted by the operator.
The UI allows one to unfold the set of impacted actionable entities to show one or many references that they contain.
When one of the references in panel~2 is selected, two things happen:
first, panel~3 shows the different recommendations the user can choose to correct the reference; and
second, panel~4 shows the source code of the actionable entity with the selected reference highlighted.
In panel~3, the ``Use this operator'' button allows one to choose a recommendation.
The user needs to accept one of the recommendations.
The ``gear and spanner'' icon (in panel~2) means that the user still needs to choose a recommendation.
The green check icon means that the user already chose a recommendation.

To prepare an evolution of the database schema, the DBA enters the list of operators he wishes to perform in our tool.
Then he selects each reference with a ``gear and spanner'' icon and chooses one of the recommendations associated to this reference.
Several concurrent recommendations might be proposed, for example when a column name changes, it can be aliased in the \ct{SELECT} clause of a view query to isolate other referencing queries from the change or it can be propagated to these other queries. 
Once all the choices are made (all operators in panel~1 have the ``green check'' icon), the DBA can click the ``Generate patch'' button (top left of the tool) to generate the SQL script.

\section{Experiments with our approach}
\label{sec:validation}

We experimented our approach on a real database in use at our university:
AppSI is a PostgreSQL database used for managing faculty members, teams, funding support, etc.
It is a proprietary database mostly developed by a single database architect.
This database is used by software systems written in different programming languages. 
This prompted the database architect to implement many client functionalities directly inside the database as stored procedures. 
At the time of the experiment, the database schema counted 95 \ct{Tables}, 62 \ct{Views}, 20 \ct{Triggers}, 64 \ct{Stored procedures} and 19 \ct{Trigger Functions}.

Because of its size and because the database schema is not open source, we cannot provide here the full details of the database model.
Figure \ref{fig:exampledb} is a tiny part of this model.

We give in the following the result of three experiments with our approach and tool on evolutions of the AppSI database.

\subsection{Replicating a non-trivial evolution}

As a preparatory step for this research, we recorded the DBA's screen during
a real migration \cite{Delp18c}.
We observed that he prepared the migration by establishing a road map in natural language listing all the actions to perform with the SQL commands to implement them.
We also observed that the initial road map was not complete nor detailed enough \cite{Delp18c}.
Following a long manual process, the architect incrementally translated his initial road map into an SQL patch to migrate from one version of the database to the next one.
He needed multiple tools to perform a trial-and-error process and find dependencies between entities of the database.
He implemented part of the patch and ran it in transactions that were always rolled back.
Whether each run of the partial patch failed or succeeded, the architect gained a better understanding of all dependencies and what he needed to do.
It took him one hour to build incrementally the SQL patch.
This patch was $\sim200$ LOC and composed of $19$ SQL statements.

Informally, we can compare the initial road map created by the DBA to the initial list of operators that we give to our tool.
The manual process to translate this into an SQL patch is what our tool helps doing, by pointing out the dependencies and offering recommendations to resolve the impacts.
Therefore, to validate our approach, we wanted to generate the same SQL patch with our tool.
This is interesting because it is a non-trivial evolution.
This experiment is also related in \cite{Delp20a}.

Based on the natural language description of the work to do in the initial road map, we extracted seven operators that the architect meant to apply for the evolution:
\begin{enumerate}
\item \ct{RenameColumn(person.uid, login)}
\item \ct{RemoveFunction(key\_for\_uid(varchar))}
\item \ct{RemoveFunction(is\_responsible\_of(int4))},
\item \ct{RemoveFunction(is\_responsible\_of(int4,int4))}
\item \ct{RenameFunction(uid(integer), login(integer))}
\item \ct{RenameLocalVariable(login.uidperson, login.loginperson)}
\item \ct{RemoveView(test\_member\_view)}
\end{enumerate}

We entered these operators in our tool (panel~1 in Figure \ref{fig:dbevolutiontool})  and let it guide us through the decision process to generate the SQL migration patch.

Fifteen decisions were taken to choose among the proposed recommendations.
They are all concerned with the renaming or aliasing of column references.
The architect told us that, as a rule, he preferred to avoid using aliases and renamed the columns, so we followed this rule in our decisions.

We finished the experiment by executing the SQL patch generated by our tool on an empty copy (no data) of the database.

To evaluate the patch, first, we checked whether it ran without errors;
second, we compared the state of the database after the architect's migration and ours.
The comparison was done on a dump of the SQL schema of both databases and comparing these two dumps with a textual diff tool.

The results of the experiment are:
\begin{enumerate}[label=\roman*]
\item The generated SQL patch was successfully applied on the database\footnote{Actually on a copy of the database without data}.

\item The diff between the two databases showed one single difference:
a comment in one stored procedure was modified in the hand-written version and not in the generated version.
We did not include operators on comments in our approach.

\item Encoding the list of operators and taking decisions took approximately
15 minutes for us who had no in-depth knowledge of the database schema.
This is 25\% of the time the architect needed in his manual process.
\end{enumerate}

These are very good results considering the architect has a much better knowledge of his database than us.

\subsection{Simple Re-modularization}

At our university, two instances of the AppSI database are running in two different departments.
One of these instances is accessed by a web application that uses a set of views of the database.
These views follow a naming convention where their name starts with a ``\texttt{web\_}'' prefix.
We will refer to these views as ``\emph{web views}''.
These views are used by no other application.

The other instance of the database is not accessed through the web application.
The database architect would like to extract the \emph{web views} in a separate namespace (\ct{SCHEMA}) that would not be part of the second instance.
That will make this second instance and possible future ones leaner and isolated from future evolutions of the \emph{web views}.

Moving the \emph{web views} in a new namespace would imply changing references to these views in the rest of the database.
The DBA of AppSI does not have this information as he is not the author of this part of the database.

There are 23 \emph{web views}, referencing 31 tables and one other view (not a \emph{web views}).
Fortunately, because these are in the default namespace, there is no need to modify the queries of the \emph{web views} after the evolution.
It happened that the \emph{web views} were referenced by no other views (not even among themselves), so the evolution ended up to be very easy, we entered the 23 \ct{MoveView} operators in our tool, and it checked that there were no impact and immediately showed the operator with a ``green check'' icon.
This gave the DBA the confidence that the evolution was harmless.
The patch generated from this evolution consisted of the 23 \ct{ALTER VIEW} SQL instructions and ran flawlessly.

Note that after modifying the database schema, the DBA still had to change the external web application to correctly reference the new view names.
This is a known problem of co-evolving a database and the application using it \cite{Meur16a}.
Our approach only considers the database itself.

In the end, this evolution was rather simple because the web views were already better isolated than the DBA feared.
It was nonetheless very useful in that the DBA did not have a clear understanding of all the possible impacts and was therefore wary of performing this evolution for fear of breaking the schema and because more urgent matters kept popping up.
After viewing the results of the experiment, the DBA decided to perform the evolution using our generated patch (after reviewing it), confident that it would be easier than expected.

\subsection{Advanced Re-modularization}

After the previous experiment, we used our model of the database schema to analyze other opportunities for ``architectural restructuring''.

As this is not the main subject of the paper, we will not describe in detail here how we identified different groups of entities (table, views, stored procedures, triggers) that pertained to different ``feature groups''  (eg: Member management, building management, mailing list, Ph.D. management, access management).
This is better explained in \cite[Chapter 4]{Delp20b}.
To summarize, (i) we manually identified feature groups of tables based on their names and columns; (ii) we added referencing entities to these initial groups based on cross-dependencies; and (iii) we validated the ``feature groups'' with the DBA.

This last experiment is similar to the previous one on a larger scale.
We want to create a namespace for the ``Ph.D. management'' feature group.
This group includes 12 tables, 
17 views, 
6 stored procedures, 
11 triggers, and 9 trigger stored procedures.

After creating the new namespace, we entered all the operators in the tool.
The good news is that many operators have no impact (because the entity is not referenced otherwise in the database schema).
Only one trigger stored procedure and two stored procedures (called by the trigger stored procedure) required choosing a recommendation to handle the impact of the \ct{Move} operator.
The tool suggested to change the name of the stored procedures in the trigger (from \ct{functionName()} to \ct{newSchema.functionName()} and we accepted the recommendation.
The generated patch ran flawlessly.

This experiment is significant as it triggers reflections on how entities are identified.
Move operators change the name of entities (see above \ct{newSchema.functionName()}).
Yet in SQL commands the name is the \emph{defacto} identifier to access entities ({eg.} \ct{ALTER TABLE xyz \ldots}).
It is therefore important to take that into account when generating the patch, if a move or rename operator is used on an entity, it impacts the following SQL commands on the same entity.

\section{Threats to Validity}

Validating tools handling the impact of software change is not easy.
On the 18 approaches reviewed by Lehnert in a meta-review \cite{Lehn11a}, only six have experimental results with metrics on the size of the system, time, precision and recall.
Only one of these has results on all the metrics together.
Having access to database schema evolutions is even more difficult.
They are less systematically recorded in version control systems and when this is the case, finding what was the original intent behind the evolution (initial operators) would require having direct access to the DBA.

Through personal contacts, we had access to the AppSI database and its DBA accepted to validate some of our experimental results.

\subsection{Internal Validity}

Internal validity is the extent to which we can draw a causal link between the treatment in the experiment and the response.

We believe there is no internal threat here because the experiment consisted in applying the approach and verifying that it could generate the correct SQL script while assisting the DBA in the process.

We verified that, in each of the three cases, the generated scripts produced the expected evolution of the database schema, and the tool did offer recommendations for the operators that required additional actions.

\subsection{External Validity}

External validity is the extent to which we can apply the findings of the study to a broader context.

There is usually a trade-off between internal and external validity.
This is the case here as we experimented with only one database in PostgreSQL.

First, we successfully imported other open-source PostgreSQL databases (ex: Liquidfeedback\footnote{\url{https://liquidfeedback.org}}) in our meta-model and we did get the tool running satisfactorily on toy evolutions.
We did not, however, have access to any real evolution on open source databases or DBA that could help us in the validation (for example by reporting past evolutions of the databases' schemas).

Second, we are aware of no specificity of the AppSI database that would suggest our tool would not work on other PostgreSQL databases.
AppSI is a real database and it is sufficiently large (95 Tables, 62 Views) to not be considered a toy experiment.

Another question is whether the approach could be applied to other RDBMS.
In this paper, we tried to clearly indicate all points where our experiment was PostgreSQL specific.
This happens mostly in the modeling part (Sections~\ref{sec:schema-modeling} and~\ref{sec:evolution-modeling}).
Because the relational database domain and the SQL language are well normalized, it seems unlikely to us that any specificities in database objects or evolution operations would appear that could not be modeled as required by our approach.

\subsection{Construct Validity}

This discusses the extent to which the results really measure what they are supposed to measure.

For the first experiment, we had the result of a real evolution performed in real life and with our approach.
We verified that the two database schemas were rigorously the same.
This was the case to the exception with a change in a comment that our tool was never intended to handle.

For the two other cases, we manually checked that the database schema was as expected after applying the generated script, and that was the case.

The only construct threat that we can imagine would be linked to a specific dependency kind in a database and/or a specific evolution operator that would create some unexpected behavior.
Other experiments would be necessary, probably with other RDBMSs to check whether this can happen.
As a research team, we do not have the manpower to develop parsers and meta-models for other RDBMSs, but we would gladly assist as best as we can anybody willing to replicate this experiment.

\subsection{Reliability}

This considers the extent to which the results can be reproduced when the research is repeated under the same conditions.

The sources for the different tools used are available on GitHub and were given in the paper.
It must be noted that we did not yet model all evolution operators and some less often used ones might be missing.
But, again, we would gladly assist anyone willing to replicate our experiment on other databases.

\section{Related Work}
\label{sec:related}

Our work may be compared to the \emph{impact analysis} and
\emph{database schema evolution} research fields.

\subsection{Impact Analysis}

The topic of Software change impact analysis was introduced by Bohnert and Arnold
\cite{Arno96b} and is now a prolific research domain.
We focus here on the much smaller domain of changes to relational databases.
Most of the work \cite{Gard06a, Kara01a, Maul08a} tries to handle the impact of database changes on client applications.
We remain fully inside the database world.

Karahasanovic and Sjøberg proposed a tool to find impacts of object-database schema changes on applications \cite{Kara01a}.
Their tool allows one to identify and visualize the impact and it also provides a language to graphically walk the impact graph.
We did not consider this idea of visualizing the impacts.

Maul \emph{et al.} \cite{Maul08a} created a static analysis technique to assess the impact of changing a relational database on its object-oriented software clients.
They use a model of the database schema but do not consider the stored procedures as we do.

Nagy \emph{et al.} \cite{Nagy10a} compared two methods for computing dependencies between
stored procedures and tables in a database:
One uses Static Execute After/Before relations \cite{Jasz08a} and the other analyzes CRUD queries and schema to find database access and propagate this dependency to stored procedures.
They did not build a full model of the procedures as we do.

Liu \emph{et al.} \cite{Liu11a, Liu13b}, proposed an ``attribute dependency graph'' to identify dependencies between columns in a database and parts of client software source code using it.

Their tool presents to the architect an overview of a change impact as a graph.

\subsection{Recommendations for Schema Evolution}

Sjøberg's work \cite{Sjob93a} studied the
evolution of a relational database and its application over 18 months.
He analyzed how many screens, actions, and queries may be affected by a potential schema change.
He does not propose recommendations to users but rather shows code locations to be manually modified.
His results suggest that change management tools are needed to handle database schema evolution.

Curino \emph{et al.} \cite{Curi08a, Curi09a} proposed PRISM, a tool suite allowing one to predict and evaluate schema modification.
PRISM also proposes a database migration feature through rewriting queries and applications to take into account the modification.
To do so, they provide a language to express schema modification operators, automatic data migration support, and documentation of changes applied to the database.

In the PRISM approach, the operators are limited to modification of structural entities of the database, whereas we also deal with change on behavioral entities.

Papastefanatos \emph{et al.} \cite{Papa08a, Papa10a} developed Hecataeus, a tool
representing the database structural entities, the queries, and the views, as a uniform
directed graph. Hecataeus allows users to create an arbitrary change and to
simulate it to predict its impact. From this perspective, it is close to the aim
of our tool. The main difference is that our
approach ensures no inconsistency is created at some point during database
evolution. It is not clear how Hecataeus addresses this problem in these papers.

Meurice \emph{et al.} \cite{Meur16a} presented a tool-supported approach that can analyze how the client source code and database schema co-evolved in the past and simulate a database change to determine client source code locations that would be affected by the change.
Additionally, the authors provide strategies (recommendations and warnings) for facing database schema change.
Their recommendations describe how to modify client program source code depending on the change performed on the database.

From the historical analysis, the authors observed that the task of manually propagating database schema change to client software is not trivial.
Some schema changes required multiple versions of the software application to be fully propagated.
Others were never fully propagated.
We argue that propagating structural changes to behavior entities of the database is a difficult task as well.

Compared to previous approaches, DBEvolution brings as a novelty that any entity can be subject to an evolution operator.
In particular, stored procedures can be modified and DBEvolution will provide recommendations for the modification.
The other way around, modifying a structural entity will provide recommendations to accommodate behavioral entities such as stored procedures or views, with the change.
This ability is absent from the above approaches.

Schuler and Kesselman [2] draw conclusions very close to ours, concerning the difficulties to manage schema evolution.
However, they work on a different problem.
Their solution is aimed at scientists storing experimental results in databases, and needing to evolve these databases' schemas.
Their focus is on structural entities and the impacts evolutions have on the data.
On the other hand, we are concerned with both structural and behavioral entities, and we do not take into account the data for the moment.
Their solution involves a Compositional High-level Schema Evolution Language.
It includes the description of schema modification operators (like us), but again their operators also deal with the data stored in the database.
Their operators are more abstract, like \emph{reify} (creates a new ``relation'', a table, from a set of columns), or \emph{domainify} (produces a domain of values from a column and its data).
Also like us, they organize the operators in a hierarchy where the root is the operator that the user needs, and the inner nodes are ``deduced'' from the root.
They have \emph{logical operators} (very abstract, expressed by the user) and \emph{physical operators} (actually executed by the RDBMS.)
We do introduce a new kind of operator (\emph{reference-oriented}) but they are not more abstract than the normal SQL operators, and the user will usually express the evolution with any of the \emph{reference oriented} or \emph{entity oriented} operators (although they would normally use the \emph{entity oriented} ones).
Finally, they have a planning step where the operators are organized in a graph and pipelined for execution.
Maybe because they target specifically data scientists, not experts in databases, rather than database architects (like us), they do not offer choices to the user (our recommendations and the \emph{HumanDecision} operator), the new operators are created from transformations that take all the decisions.
So whereas both approaches present similarities on the surface, there are profound differences in the details.
We are however interested in being able to handle consequences on the data stored like Schuler and Kesselman do, and in having higher-level operators like theirs or others.

\section{Conclusion}
\label{sec:conclusion}

We developed an approach to manage relational database schema evolution.
This approach addresses the two main constraints that an RDBMS sets:
no schema inconsistency is allowed during the evolution, and stored procedure bodies are not described by meta-data.
It also computes the impact of changes and has strategies to resolve these impacts.

Our approach makes the following contributions:
(i) a meta-model for relational databases easing the computation of the impact of a change;
(ii) models of change operators that allow to compute the impact of these operators;
(iii) a semi-automatic approach to evolve a database schema while managing the impact of the changes, and
(iv) an experiment to assess that our approach can reproduce a change that happened on a database used by a real project with a gain of 75\% of the time.

Our overall goal is to introduce database schema evolution tools that are considered normal in software engineering.
This could include quality rule checking \cite{Delp17a}, refactorings (high-level evolution operators), or testing.

An interesting extension for the usability of the tool would be to help the user decide which recommendations to choose when several are available. For example, to manage the consequences of a change, it would be interesting to order these recommendations, \emph{e.g.} according to the most likely choice of the user.

One interesting extension that we would like to develop would be to add more abstract operators to our tool, such as \emph{splitting a table in two}, or \emph{historize column} which will modify the database schema to keep track of the history of the values of a column through the database life.
Some of these abstract operators may need to deal explicitly with the data in the tables (dump and reload the data).
We believe our models should be able to handle such cases.

Our experiments involve only one database, obviously, more experiments are needed on other databases.
Also, for now, we support only PostgreSQL RDBMS.
Although SQL is a very standardized domain, we want to consider other RDBMS to verify what (if any) additional operators would be needed.

Finally, some operators require to transform or move data stored in the database.
We would like to support such operators by generating CRUD queries in the final patch.


\bibliographystyle{sn-mathphys}
\bibliography{rmod,others,local}

\end{document}